\documentstyle[12pt]{article}
\date{}
\begin{document}
\title{The $\pi \sigma$-Axial Exchange Current}

\author{K. Tsushima$^{1,3}$  and D.O. Riska$^2$}
\maketitle
\centerline{$^1$ \it Institute for Theoretical Physics, University of
Tuebingen, D72076 Tuebingen, Germany}

\centerline{$^2$ \it Department of Physics, SF-00014 University of
Helsinki, Finland}

\vspace{1cm}

\centerline{\bf Abstract}
\vspace{0.5cm}
The axial exchange current operator that arises from the coupling of
the axial field to the pion and effective scalar field in nuclei is
constructed. The spatial dependence of this exchange current operator
may be determined directly from the nucleon-nucleon interaction. This
"$\pi\sigma$"-axial exchange current contributes about 5\% to the
quenching of the nucleon axial current coupling constant $g_A^{GT}$ in
heavy nuclei.

\vspace {3 cm}
$^3$Supported in part by DFG contract Fa 67/14-1
\newpage
The value of the effective axial current coupling constant $g_A^{GT}$
of nucleons in nuclei is considerably smaller than the corresponding
free nucleon value [1,2]. Shell model analyses [3] as well as studies
of giant Gamow-Teller resonances [4] indicate that it is quenched by
about 30\% already in $sd$-shell nuclei. This quenching effect is
conventionally ascribed to the effect of $\Delta_{33}$-excitations
[5,6] and core-polarization [1]. The short range exchange currents
that are associated with the axial current coupling through
intermediate nucleon-antinucleon pairs to the short range parts of the
nucleon-nucleon interaction have been found to be insignificant [7].\\

The absolute magnitude of both the corrections that are associated
with intermediate $\Delta_{33}$-excitation mechanisms and the second
order configuration mechanisms remain somewhat uncertain as they are
sensitive to the short range components and especially to the strength of
the nucleon-nucleon tensor interaction. The large core polarization
correction obtained with a strong model for the nucleon-nucleon tensor
interaction in ref. [1] appears to explain most of the empirically
found quenching of $g_A^{GT}$ by itself. We here point out that the
axial exchange current that arises from the coupling of the axial
field to the pion and the effective scalar field in nuclei ("the
$\sigma$-meson") represents another mechanism that contributes a small
but still significant
amount to the quenching of the value for $g_A^{GT}$ in nuclei (Fig. 1). We
demonstrate that the corresponding axial exchange current operator
can be constructed directly from the
model for the nucleon-nucleon interaction, and thus involves no
parameters that do not appear in the interaction model. Using
realistic phenomenological models for the interaction we estimate that
the $\pi\sigma$-mechanism alone leads to a quenching of about 5\% of
$g_A^{GT}$. The presence of this additional quenching mechanism
suggests that the previous estimates for the core polarization
correction in ref. [1] need to be revised downwards by
employment of a weaker model for the tensor force.\\

It proves convenient to base the derivation of the effective
$\pi\sigma$ axial exchange current on the modified $\sigma$-model
[8,9]

$${\cal L}={\cal L}_0+{\cal L}_1,\eqno(1)$$
where ${\cal L}_0$ is the Lagrangian density for the usual linear
$\sigma$-model

$${\cal L}_0=\bar \psi[i\gamma^\mu \partial_\mu-g(\sigma +i\vec \tau
\cdot \vec \pi \gamma_5)]\psi$$
$$+{1\over 2}[(\partial_\mu \vec \pi)^2+(\partial_\mu
\sigma)^2]+{1\over 2} m_0^2(\vec \pi^2+\sigma^2) - {1\over
4}\lambda^2(\vec \pi^2+\sigma^2)^2,\eqno(2)$$
and ${\cal L}_1$ is the chirally symmetric term

$${\cal L}_1=C\{\bar \psi \gamma_\mu{\vec \tau \over 2}\psi\cdot (\vec \pi
\times \partial^\mu \vec \pi)+\bar \psi \gamma_\mu \gamma_5 {\vec \tau
\over 2}\psi\cdot [\vec \pi \partial^\mu \sigma-\sigma \partial^\mu
\vec \pi]\},\eqno(3)$$
which is introduced so as to allow $g_A$ to differ from $1$ at the
tree level. We here treat the $\sigma$-field as an effective
representation of the isoscalar component of the interacting $\pi
\pi$ system.\\

The Lagrangian (1) implies that the vacuum breaks chiral symmetry
spontaneously $(m_0^2>0)$. In the vacuum $<\sigma>=f_\pi$ (the pion
decay constant). The
dynamical isoscalar $\sigma$-field is defined as
$\sigma'=\sigma-f_\pi$. We here view the field $\sigma'$ as an
effective representation of the isoscalar part of the interacting
$\pi \pi$ system. \\

The parameters $g$ and $C$ in (2) and (3) are determined as [9]

$$g_A=1+C f_\pi^2=1.262,\eqno(4a)$$
$$g=G/g_A,\eqno(4b)$$
where $G$ is the usual $\pi NN$ (pseudoscalar) coupling constant
$(\approx 13.5)$. \\

The effective $\pi NN$ and $\sigma NN$ couplings that are implied by the
Lagrangian model (1) are

$${\cal L}_{\pi NN}\simeq -iG\bar \psi \vec \tau \cdot \vec \pi
\gamma_5 \psi,\eqno(5a)$$
$${\cal L}_{\sigma NN}\simeq -\frac{G}{g_A}\bar \psi
\sigma'\psi.\eqno(5b)$$
The $\pi \sigma$ axial vector coupling is determined by the axial
current operator [9]:

$$A_\mu^i=g_A\bar \psi \gamma_\mu \gamma_5 {\tau^i \over 2}\psi
+\pi^i\partial_\mu \sigma'-\sigma'\partial_\mu \pi^i
-f_\pi \partial_\mu \pi^i.\eqno(6)$$

Using these expressions the $\pi \sigma$-axial exchange current
operator is found to be

$$\vec A_i=\frac{G^2}{m_Ng_A}(\vec k_1-\vec k_2)\{\frac{\vec
\sigma^2\cdot \vec k_2 \tau_i^2}{(k_2^2+m_\pi^2)(k_1^2+m_\sigma^2)}$$
$$-\frac{\vec \sigma^1 \cdot \vec
k_1\tau_i^1}{(k_1^2+m_\pi^2)(k_2^2+m_\sigma^2)}\}.\eqno(7)$$
Here $m_N$ is the nucleon, $m_\pi$ the pion and $m_\sigma$ the mass of
the effective scalar meson. The momentum variables $\vec k_1$ and
$\vec k_2$ denote the momentum transfers to nucleons $1$ and $2$. Note
that the corresponding charge operator is proportional to the
difference between the energies of the two mesons, both of which are
small, and hence that operator is insignificant in comparison to the
axial exchange charge operators associated with both pion [10] and short
range exchange [11] mechanisms.\\

To assess the importance of the $\pi \sigma$-exchange current (7) we
shall reduce it to the form of a correction to the effective single
nucleon axial current operator:

$$\vec A^i=-g_A(1-\delta_A)\vec\sigma {\tau^i \over
2}-\frac{g_P(1-\delta_P)}{2m_Nm_\mu}\vec q\vec \sigma \cdot \vec
q{\tau^i \over 2}.\eqno(8)$$
Here $m_\mu$ is the muon mass.
The terms 1 in the factors $(1-\delta_A)$ and $(1-\delta_P)$
represent the current of a free nucleon and the terms $\delta_A$ and
$\delta_P$ represent the exchange current contribution. Note that the
induced pseudoscalar coupling constant $g_P$ is related to $g_A$ as

$$g_P=-\frac{2m_Nm_\mu}{q^2+m_\pi^2}g_A\eqno(9)$$
for free nucleons.\\

The reduction of the $\pi\sigma$-axial exchange current to an
effective single nucleon current operator leads to a direct term and
an exchange term. Since $\vec q=\vec k_1+\vec k_2$ the direct term is
readily found to contribute only to the induced pseudoscalar current,
the magnitude of the contribution being

$$\delta_P=\frac{G^2}{g_A^2m_Nm_\sigma^2}\rho,\eqno(10)$$
where $\rho$ is the nucleon density. This result was first obtained in
ref. [9] as the nuclear renormalization correction to the pion decay constant.
The other mechanisms that contribute to $\delta_P$ have been
discussed recently in ref. [12].\\

The exchange term leads to a contribution to $\delta_A$. Taking $\vec
q=0$ (i.e. $\vec k_1=-\vec k_2$) we find, using the Fermi gas model,
this contribution to be

$$\delta_A=\frac{G^2}{g_A^2m_Nm_\sigma^2}\rho\; I,\eqno(11)$$
where $I$ is the integral

$$I=4x_\sigma^2\int_{0}^{1}dx\frac{x^4(1-x)}{(x^2+x_\pi^2)(x^2+x_\sigma^2)}$$

$$=2x_\sigma^2+\frac{4x_\sigma^2}{x_\sigma^2-x_\pi^2}\{x_\pi^3 arc\,
tan {1\over x_\pi}-x_\sigma^3 arc\, tan {1\over x_\sigma}$$
$$-\frac{x_\pi^4}{2}log (1+\frac{1}{x_\pi^2})+\frac{x_\sigma^4}{2}log
(1+\frac{1}{x_\sigma^2})\}.\eqno(12)$$
with

$$x_\pi\equiv m_\pi/2k_F,\, x_\sigma\equiv m_\sigma/2k_F,\eqno(13)$$
where $k_F$ is the Fermi momentum. In the limit of a zero pion and an
infinite $\sigma$-meson mass the integral $I$ takes the value $1/3$.
To obtain an estimate for the first factor in (11) we take $m_\sigma
\approx m_N$ and $\rho=2k_F^3/3\pi^2$, with $k_F=1.40 fm^{-1}$, which
is the value for normal nuclear matter. This qualitative estimate of
the $\pi\sigma$ contribution to the quenching of $g_A$ yields the
value $\delta_A\simeq 0.066$. This value is quite close to the value
calculated below using a realistic boson exchange potential model and
taking into account the corresponding short range correlations.\\

In order to reduce the model dependence of the $\pi\sigma$ axial
exchange current operator (7) we note that it can be expressed in
terms of the effective isospin independent scalar $v_S^{+}$ and
isospin dependent pseudoscalar $v_P^{-}$ components of the
nucleon-nucleon interaction as

$$\vec A_i=-\frac{g_A}{m_NG^2}(\vec k_1-\vec k_2)\{v_S^{+}(\vec
k_1)v_P^{-}(\vec k_2)\vec \sigma^2\cdot \vec k_2 \tau_i^2$$
$$-v_S^{+}(\vec k_2)v_P^{-}(\vec k_1)\vec \sigma^1 \cdot \vec
k_1\tau_i^1\}.\eqno(13)$$
Here $v_S^{+}(\vec k)$ and $v_P^{-}(\vec k)$ are the Fourier transforms
of the scalar and pseudoscalar potential components of the
nucleon-nucleon interaction defined as in
[13]. Here we have replaced the factors $G^2/(k^2+m_\pi^2)$ in (7) with
$v_P^{-}(k)$ and the factors $(k^2+m_\sigma^2)^{-1}$ by $-g_A^2
v_S^{+}(k)/G^2$. This interpretation of the meson propagators in (7)
as the corresponding components of the nucleon-nucleon interaction is
natural and allows us to determine the short range behaviour of the
exchange current operator directly from the nucleon-nucleon model
without the need to introduce ad hoc cut off parameters.
With this expression for the $\pi\sigma$-exchange current the
contribution to the quenching factor $\delta_A$ (11) is replaced by
the expression

$$\delta_A=-\frac{16k_F^2}{m_NG^2}\rho\int_{0}^{1}dxx^4(1-x)v_P^{-}
(2k_Fx)v_S^{+} (2k_Fx).\eqno(14)$$
Note that all terms in this expression with the exception of $m_N$ and
$k_F$ are completely determined by the
nucleon-nucleon interaction model.\\

If we construct the potential components $v_P^{-}$ and $v_S^{+}$ using
the parametrized Paris potential [14] we obtain the value $-3.60
fm^{4}$ for the integral in (14) and correspondingly at nuclear
matter density the value $\delta_A=0.024$. Using the Bonn potential model [15]
we find the value $-8.98 fm^{4}$ for the integral in (14) and thus the
value $\delta_A=0.060$ at nuclear matter density. These values are
further reduced somewhat by short range correlations: if the spin
averaged correlation function that is obtained by constructing the
$G$-matrix for these two potential models is taken into account in the
integrals in (14) the numerical values for $\delta_A$ are reduced to
$\delta_A=0.023$ (Paris) and $0.055$ (Bonn) respectively. \\

The very small value for $\delta_A$ that is obtained with the Paris
potential model is a consequence of the anomalous repulsive behaviour
of the effective isospin independent scalar potential component
at short distances
in that model [13]. We therefore view the larger value obtained with
the Bonn potential to be the more realistic one. The conclusion is then that
the effective $\pi\sigma$ axial exchange current causes a $\approx
5\%$ quenching of the axial current coupling constant $g_A^{GT}$ at
nuclear matter densities. This
is somewhat larger than the corresponding contribution to the
quenching caused by the $\pi\rho$ exchange mechanism [17].\\

If the 5\% contribution of the $\pi\sigma$-axial exchange current to
the quenching of $g_A^{GT}$ is combined with the 20-23\% quenching due
to the $\Delta_{33}$ excitation, $\pi\rho$ exchange and short range
exchange currents obtained in ref. [7] the predicted net exchange
current caused quenching of $g_A^{GT}$ becomes 25-28\% at nuclear
matter density. If this number is combined with the large quenching
(40\%) estimated to arise from second order configuration mixing in
ref. [1] the predicted quenching of $g_A^{GT}$ is about 2 times too
large. As the latter effect was calculated in ref. [1] on the basis of
the Hamada-Johnston potential [18], which has a very strong isospin
dependent tensor component, it would be important that this calculation
be redone with a modern weaker tensor potential in order to obtain a
quantitative estimate for a weaker core polarization contribution.
Another reason to believe that the second order configuration mixing
contribution to $g_A^{GT}$ in ref. [1] may be unrealistically large is
the fact that the effective tensor force in the medium is also
somewhat weakened by a
core polarization induced screening mechanism [19].\\

The $\pi\sigma$ axial exchange current operator (7), (13) will also
contribute to the exchange current correction to the transition rate
for $\beta$-decay of $^3H$. To obtain an estimate for the magnitude of
this contribution one may compare the expression (7) for the
$\pi\sigma$ axial exchange current operator to the corresponding one
for the $\pi\rho$ axial exchange current [20]:
\newpage
$$\vec A_i(\pi\rho)=-g_A\frac{g_\rho^2}{m_N}\frac{\vec \sigma^2\cdot
\vec k_2}{(m_\rho^2+k_1^2)(m_\pi^2+k_2^2)}(\vec \tau^1 \times \vec
\tau^2)_i.$$
$$\{(1+\kappa)\vec \sigma^1 \times \vec k_1-i(\vec p_1+\vec
p'_1)]+(1\leftrightarrow 2).\eqno(15)$$
Here $g_\rho$ is the $\rho NN$ vector coupling constant, $\kappa$ the
$\rho NN$ tensor coupling and $m_\rho$ is the $\rho$-meson mass. The
symbol $(1\leftrightarrow 2)$ stands for a term in which all particle
coordinates are exchanged. The contribution of this operator to the
$\beta$-decay rate of $^3H$ has been found to be only about $0.7\%$ in
a calculation based on several realistic wavefunctions [20].\\

The local part of the $\pi\rho$ axial exchange current operator is
very similar in form to that of the $\pi\sigma$ axial exchange current
operator (7). Moreover the numerical values of the coefficients in
front are also of similar order of magnitude: $G^2/g_A\simeq 145$
versus
$g_A g_\rho^2(1+\kappa)\simeq 60$ (with $g_\rho^2/4\pi=0.5\,,
\kappa=6.6$). As, however, only the component of (7) that is
antisymmetric in isospin contributes to the $\beta$-decay of $^3H$ the
factor 145 above should be divided by 2 before the comparison, and thus
the numerical coefficients in the $\pi\sigma$ and $\sigma\rho$ axial
exchange current operators differ by only about 20\%. Finally the
mass of the "effective" $\sigma$-meson in the nucleon-nucleon
interaction is close to that of the $\rho$-meson. Although the
matrix elements of the two operators between the $^3H$ and $^3He$ are
not identical it is nevertheless evident from this comparison that the
contribution of the $\pi\sigma$ axial exchange current to the rate
for $\beta$-decay of $^3H$
cannot be much larger than that of the $\pi\rho$ axial exchange
current (which is very small). A quantitative calculation of that
reaction would nevertheless be important, as this mechanism will also
contribute to the solar neutrino reactions $pp\rightarrow
d+e^{+}+\nu_e$ and $p+ ^3He\rightarrow$ $^4He+e^{+}+\nu_e$. As in the
latter reaction the $\pi\rho$ exchange current operator has found to
be significant, we expect the $\pi\sigma$ axial exchange current
operator to be of equal significance [20].

\newpage
\begin{enumerate}
\item A. Arima et al., Adv. Nucl. Phys. {\bf 18} (1987) 1
\item F. Osterfeld, Rev. Mod. Phys. {\bf 64} (1992) 491
\item B.A. Brown and B.H. Wildenthal, Phys. Rev. {\bf C28} (1983) 2397
\item C. Gaarde et al., Nucl. Phys. {\bf A334} (1980) 24
\item M. Rho, Nucl. Phys. {\bf A231} (1974) 493
\item K. Ohta and M. Wakamatsu, Nucl. Phys. {\bf A234} (1974) 445
\item K. Tsushima and D.O. Riska, Nucl. Phys. {\bf A549} (1992) 313
\item B.W. Lee, Chiral Dynamics, Gordon and Breach, New York (1972)
\item E.Kh. Akhmedov, Nucl. Phys. {\bf A500} (1989) 596
\item K. Kubodera, J. Delorme and M. Rho, Phys. Rev. Lett. {\bf 40}
(1978) 755
\item M. Kirchbach, D.O. Riska and K. Tsushima, Nucl. Phys. {\bf A542}
(1992) 616
\item M. Kirchbach and D.O. Riska, The Effective Induced Pseudoscalar
Coupling Constant, Preprint HU-TFT-93-46 (1993)
\item P.G. Blunden and D.O. Riska, Nucl. Phys. {\bf A536} (1992) 697
\item M. Lacombe et al., Phys. Rev. {\bf C21} (1980) 861
\item R. Machleidt, K. Holinde and Ch. Elster, Phys. Repts {\bf 149}
(1987) 1
\item M. Hjorth-Jensen, M. Kirchbach, D.O. Riska and K. Tsushima,
 Nucl. Phys. {\bf A563} (1993) 525
\item K. Tsushima, D.O. Riska and P.G. Blunden, Nucl. Phys. {\bf A559}
(1993) 543
\item T. Hamada and I.D. Johnston, Nucl. Phys. {\bf 34} (1962) 382
\item K. Nakayama, Phys. Lett. {\bf 165B} (1985) 239
\item J. Carlson, D.O. Riska, R. Schiavilla and R.-B. Wiringa, Phys.
Rev. {\bf C44} (1991) 619
\end{enumerate}

\vspace{1cm}

{\bf Figure Caption}
\vspace{0.5cm}

Fig. 1.$\quad$ Diagrammatic representation of the $\pi\sigma$
axial exchange current operator

\end{document}